OPINION

# On Boltzmann's genius and thermodynamics

Elias P. Gyftopoulos
Massachusetts Institute of Technology
77 Massachusetts Avenue; Room 24-111
Cambridge, MA  02139  USA

A recent essay [1] reminds us of how richly Boltzmann deserves to be admiringly commemorated for the originality of his ideas on the occasion of his 150$^{th}$ birthday.  Without any doubt, the scientific community owes Boltzmann a great debt of gratitude for his ingenious and pathfinding contributions.  However, the essay chooses to illustrate this important memorial by statements and inferences that perhaps are questionable today even to Boltzmann himself.  I will comment only on three issues.

MICROSCOPIC VERSUS MACROSCOPIC BEHAVIOR

Schrödinger is quoted as saying: "Boltzmann's ideas really give an understanding of macroscopic behavior", but another remark by him [2] is not included: "The older and more naïve application is to N actually existing physical systems in actual physical interaction with each other, e.g., gas molecules or electrons…" … "This original point of view is associated with the names of Maxwell and Boltzmann, and others.  But it suffices only for dealing with a very restricted class of physical systems – virtually only with gases.  It is not applicable to a system which does not consist of a great number of identical constituents with private energies." … "Hence a second point of view…, which we owe to Willard Gibbs, has been developed."

In fact, the latter comment by Schrödinger is even more restrictive than he intended.  If the Boltzmann equation [3, 4] is valid, careful scrutiny of its terms reveals that it is applicable only if the value (expectation value) of the number of molecules of the gas in each elementary volume of the phase space is smaller than unity [5].  To see this clearly, we recall that the Boltzmann equation is given by the relation

$$\frac{\partial f}{\partial t} + \nabla_r \cdot \mathbf{v} f + \nabla_v \cdot \mathbf{a} f = \text{Collision integral} \qquad (1)$$

In the left-hand side of the equation, $f(\mathbf{r}, \mathbf{v}, t) d\mathbf{r} d\mathbf{v}$ represents the (expected) number of molecules of a system in phase-space volume $d\mathbf{r}d\mathbf{v}$, and the normalized sum of the three terms is the time derivative of this number, $(df/dt)d\mathbf{r}d\mathbf{v}$.  In the collision integral, $f(\mathbf{r}, \mathbf{v}, t)d\mathbf{r}d\mathbf{v}$ enters as the probability that upon measurement at time t, a molecule of the system within the volume [6] between $\mathbf{r}$ and $\mathbf{r} + d\mathbf{r}$ is found to have velocity between $\mathbf{v}$ and $\mathbf{v} + d\mathbf{v}$.  The circumstance for which a number is both a number of molecules and a probability occurs only if the number is smaller than unity.  So Boltzmann's equation and the resulting H-theorem [7] are applicable not to macroscopic systems with very large numbers of molecules but to systems so dilute as to have an expected number of molecules less than unity. This observation is a tribute to the genius of Boltzmann who, unknowingly, presaged the coming of quantum theory.

An excellent application of Boltzmann's equation is to neutrons in a nuclear power reactor. At full power, the number of neutrons is less than $10^{18}\,\#/m^3$ or less than unity per cube of a side of $10^{-6}\,m$. This side is huge compared to the neutron diameter (about $10^{-15}\,m$).

REVERSIBILITY AND THE AGE OF THE UNIVERSE

The essay states [1]: "We can expect to see unusual events such as gases unmixing themselves, only if we wait for times inconceivably long compared with the age of the universe". This statement has its origin in a calculation made by Boltzmann [8] in response to objections to the statistical interpretation of irreversibility raised by Poincaré [9] and Zermelo [10]. However, it perpetuates a fallacy that has plagued the debate about statistical mechanics for over a century because it overlooks both the thermodynamic definition of reversibility, and fundamental results of quantum theory.

Even if valid, the inconceivably long time calculated by Boltzmann describes only one reversible process, the spontaneous return to initial conditions via intermolecular collisions [8]. However, such a calculation ignores the thermodynamic definition of reversibility. According to this definition, a process is reversible if both the system and its environment can be restored to their respective initial states [11-13] along any path of states and not exclusively along the path of intermolecular collisions. Moreover, a process is irreversible if, upon restoring the system to its initial state, an irreducible permanent effect is left on the environment. The permanent effect is experienced even if the restoration of the system is achieved with perfect machinery.

In light of the definitions just cited, the age of the universe is interesting but not germane to the issue at hand. Three universally and daily observed examples illustrate this conclusion.

(a) Initially, a well insulated bucket of water consists of half hot and half cold water. If they are allowed to interact only with each other, the two water halves reach mutual stable equilibrium (thermodynamic equilibrium), all the water is lukewarm and, of course, the process experienced by the water is irreversible. It is irreversible because we can always restore the hot and cold water parts over a very short period of time by means of cyclic machinery, but such restoration leaves a permanent effect on the environment, increases the entropy of the environment, even if the machinery is thermodynamically perfect.
(b) Initially, a high-quality, electricity storage, charged battery is wrapped in excellent insulation and left idle on a shelf. After a few years, the battery is found to be dead because of internal discharge at constant energy. At that time, we can restore the initial state of the battery (by first cooling and then recharging) over a short period of time, much shorter than the time required for the completion of the spontaneous internal discharge. The spontaneous discharge is irreversible because the restoration to the charged state increases the entropy of the environment even if the cooling and the recharging processes are perfect.

(c) A third example is the chemical reaction $2H_2 + O_2 = 2H_2O$ in an isolated combustion chamber. The reaction is irreversible but the electrolytic reformation of $H_2$ and $O_2$ can be accomplished very quickly.

In the three examples, plus innumerable other irreversible processes, there is practically no difficulty in restoring the initial state of the system over a period of time that is short or long but has no connection to the age of the universe. Moreover, regardless of the time interval, each restoration consumes no energy (principle of energy conservation) but dissipates natural resources (coal, oil, natural gas, or solar radiation), that is, materials with the same energy but higher entropy than the corresponding initial values are accumulated in the environment. This accumulation occurs even if the restoration devices are perfect.

Apart from the experiences, there is a fundamental theoretical objection to both the validity of the inconceivably long time calculated by Boltzmann, and the recurrence of the initial state via velocity reversals [14]. The objection arises from a basic result of quantum theory, has no classical analogue, and is very often overlooked. It asserts that, in thermodynamic equilibrium, the value of the velocity of each and every molecule is zero. Accordingly, neither the collisions invoked by Boltzmann [8] nor velocity reversals [9, 10] can restore the initial state over either a short or a long period of time.

In support of the assertions just cited, and to simplify the algebra, I consider a system A having a value of the amount of each constituent equal to an eigenvalue of the corresponding number operator, a fixed volume, and nondegenerate energy eigenstates. The most general description of the thermodynamic or stable equilibrium states of A is quantum-theoretic not only because classical mechanics is a special case of quantum physics but, more importantly, because there exist quantal aspects which can be neither represented classically nor overlooked. For example, the question of stability of molecular energy levels cannot be answered by means of classical mechanics.

As it is very well known, the energy E, entropy S, and quantum-mechanical canonical properties $x_i$ associated with a thermodynamic equilibrium state of the system under consideration are given by the relations

$$E = \sum_i x_i \epsilon_i ; \qquad S = -k \sum_i x_i \ln x_i ; \qquad x_i = \frac{\exp(-\beta \epsilon_i)}{\sum_i \exp(-\beta \epsilon_i)} \qquad (2)$$

where $\epsilon_i$ is the i-th energy eigenvalue, k the Boltzmann constant, $\beta$ a coefficient determined solely by the value of the energy E [15], and the energy eigenvalues satisfy the eigenvalue problem

$$\hat{H} \psi_j = \epsilon_j \psi_j \qquad (3)$$

Regardless of the number of molecules of the constituents of the system, any position operator $\hat{x}_k$ and the momentum operator $\hat{p}_k$ along the direction of the coordinate $x_k$ of a molecule of mass m satisfy the commutator relation

$$\left[\hat{x}_k, \hat{H}\right] = i\frac{\hbar}{m}\hat{p}_k \tag{4}$$

and the Heisenberg uncertainty relation [16]

$$\Delta x_k \Delta E \geq \frac{\hbar}{2m}\left|<\hat{p}_k>\right| \tag{5}$$

where $\Delta x_k$ and $\Delta E$ are the standard deviations (fluctuations) of measurement results of the observables $\hat{x}_k$ and $\hat{H}$, respectively.

For any stationary energy eigenstate $\psi_j$ of a system with finite dimensions

$$\Delta x_k < \infty \tag{6}$$

$$(\Delta E)^2 = <\psi_j^* \hat{H}^2 \psi_j> - <\psi_j^* \hat{H} \psi_j>^2 = \epsilon_j^2 - \epsilon_j^2 = 0 \tag{7}$$

and, therefore, relation (5) yields $\left|<\hat{p}_k>\right| = 0$. It follows that each stationary energy eigenstate corresponds to a zero value of momentum of any molecule in any direction, and that the same must be true for a canonical combination of such eigenstates. Accordingly, starting from a thermodynamic equilibrium state, neither collisions nor velocity reversals can restore the system to its initial state over a short or a very long period of time because all molecules are at a standstill. Said differently, even an omnipotent Maxwellian demon cannot accomplish his task because there are no fast and slow molecules to be sorted out.

This result cannot and should not be regarded as a criticism of Boltzmann's genius because quantum-theoretic ideas were unknown in the 19th century.

EVEN ARISTOTLE(!) WAS NOT ALWAYS RIGHT

The essay further states [1]: "…all claims of inconsistencies that I know of, in my opinion, are wrong; I see no need for alternate explanations." In view of the historical evolution of the ideas of physical sciences, the summary dismissal for all claims of inconsistencies as wrong, and the blanket denial of even the possibility of alternate explanations are very puzzling. To stimulate debate, in addition to the preceding arguments I would like to suggest thorough review of the following evidence. Over the past two decades a small group of physicists and engineers had the privilege and good fortune to undertake and complete the following programs: (a) state the principles and implications of general thermodynamics without ambiguities, inconsistencies, and circular arguments, and without any reference to inherent or statistical probabilities [17]; thus, not only entropy is found to be a property of both microscopic and macroscopic matter in both equilibrium and nonequilibrium states, but also the purely thermodynamic characteristics of entropy (such as additivity) are established; (b) observe that the von Neumann concept of a homogeneous ensemble that represents a projector (every member of the ensemble is described by the same projector or wave function as any other member) can be readily extended to density operators (every member of the ensemble is described by the same density operator $\rho$ as any other member, that is, the ensemble is not a mixture of projectors); this extension is accomplished without any fundamental changes of the quantum-theoretic postulates about observables, measurement results, and values of observables [18, 19]; thus the monstrosity of the concept of state that concerned Schrödinger [20] and Park [21] is eliminated; (c) using the thermodynamic characteristics in part (a), establish criteria that must be satisfied by any analytical expression that purports to represent the entropy of thermodynamics [22, 23]; thus, prove that Boltzmann's entropy as well as many other expressions in the literature fail the criteria, and that the only acceptable expression is $-kTr\rho\ln\rho$ for the $\rho$'s specified in part (b); (d) establish that the quantum-theoretic $\rho$'s specified in part (b) allow the unification of quantum theory and thermodynamics without any need for statistical (subjective or informational) probabilities [24, 25]; (e) prove that the unified theory applies to all systems, small or large (including a one-spin system), and all states, nonequilibrium, equilibrium, and stable (or thermodynamic) equilibrium [24]; (f) recognize that Schrödinger's equation of motion (or its von Neumann equivalent) does not describe reversible but nonunitary transformations of state [19, 24]; thus disclose the need for a complete equation of motion that describes all reversible (unitary and nonunitary) and all irreversible processes, and discover (not derive!) such an equation [26, 27]; the equation cannot be derived because it is a postulate and not a theorem, as $F = ma$ is a postulate and not a theorem; (g) establish criteria that must be satisfied by any equation of motion that purports to be the complete equation of motion of physics [23, 28]; and (h) prove that of all the equations of motion that have appeared in the literature [29], and that claim to regularize more phenomena than those encompassed by Schrödinger's equation of motion, the only one that satisfies all the criteria in part (f) is the equation presented in Refs. 26 and 27.

Because the programs just cited have been successfully carried out to a very high degree of completion, I know of no compelling a priori reason, either experimental or theoretical, to deny their rigor and to reject their need. On the contrary, because they may be more widely proven to be important contributions to and novel advances of scientific thought, they should be rigorously reviewed and rejected only if they are found faulty. The rejection, however, cannot and should not be based solely on the fact that the new paradigm differs from the old one because if it did not differ it would not be new. Rather, the evaluation should consider the scientific merits and demerits of what is being proposed.

It seems to me that such a review would be a most fitting tribute to the genius of Boltzmann, who after all became famous not by extrapolating from the past but by daring to be creative and unorthodox.